\documentclass[twocolumn,twocolappendix]{aastex631}
\usepackage{CJK}

\usepackage{amsmath}
\usepackage{appendix}

\DeclareGraphicsExtensions{.eps,.eps.gz,.epsi}

\shorttitle{Gaia BP/RP spectra spatial uniformity check}
\shortauthors{Huang et al.}

\graphicspath{{./}{my_figures/}}

\begin{document}
\begin{CJK*}{UTF8}{gbsn}
\title{A Spatial Uniformity Check of {\it Gaia} DR3 Photometry and BP/RP Spectra}

\author[0000-0002-1259-0517]{Bowen Huang (黄博闻)}
\affiliation{Institute for Frontiers in Astronomy and Astrophysics, Beijing Normal University, Beijing, 102206, China}
\affiliation{School of Physics and Astronomy, Beijing Normal University No.19, Xinjiekouwai St, Haidian District, Beijing, 100875, China}

\author[0000-0003-2471-2363]{Haibo Yuan (苑海波)}
\affiliation{Institute for Frontiers in Astronomy and Astrophysics, Beijing Normal University, Beijing, 102206, China}
\affiliation{School of Physics and Astronomy, Beijing Normal University 
No.19, Xinjiekouwai St,
Haidian District, Beijing, 100875, China}

\author[0000-0001-8424-1079]{Kai Xiao (肖凯)} 
\affiliation{School of Astronomy and Space Science, University of Chinese Academy of Sciences, Beijing, 100049, China}

\correspondingauthor{Yuan Haibo (苑海波)}
\email{yuanhb@bnu.edu.cn}

\begin{abstract} 
Gaia DR3 photometry and BP/RP spectra have been widely used as reference in photometric calibrations. In this work,
we check the spatial uniformity of Gaia DR3 photometry and BP/RP spectra by comparing the  $BP$, $RP$ and $G$ bands photometry with the synthetic ones from the BP/RP spectra. The discrepancies have a 
small dispersion of 1.07, 0.55 and 1.02 mmag for the $BP$, $RP$, and $G$ bands, respectively.
However, the discrepancies exhibit obvious spatial patterns, which are clearly associated with the Gaia's scanning law.
The patterns observed in the $BP$ and $G$ bands are similar, with discrepancies between photometry and spectra being more pronounced in these bands compared to the $RP$ band.
A further independent test with the Dark Energy Survey DR2 photometry reveals that the spatial patterns are more likely attributed to the Gaia DR3 BP/RP spectra, particularly in the $BP$ band.
On one hand, our results confirm the high spatial uniformity of Gaia data at the mmag level. On the other hand, 
our results suggest that the spatial uniformity of Gaia DR3 BP/RP spectra is not as good as that of Gaia DR3 photometry, and could be further improved in the future.
\end{abstract}


\section{Introduction} 
Gaia Data Release 3 (DR3; \citealt{dr3content}) has made an extensive and novel data-set available, including the $BP$, $RP$ and $G$ bands photometry for over 1.8 billion sources and 220 million low-resolution BP/RP (XP; hereafter) spectra. Despite of modest magnitude- and color-dependent calibration errors (e.g., \citealt{NZXGaiaColor}; \citealt{YangGaiaMag}; \citealt{HuangXPcor}), Gaia DR3 is acknowledged for its superior spatial uniformity at the mmag level， thanks to its survey and calibration strategies. 
Both Gaia DR2/DR3 photometry and XP spectra, after correction of magnitude-dependent calibration errors in most cases, are extensively employed as standard stars for photometric calibrations of various surveys.
Examples include 
the Sloan Digital Sky Survey Stripe 82 (\citealt{S82_2021};\citealt{S82}), 
the Stellar Abundance and Galactic Evolution Survey (\citealt{SAGES};  \citealt{SAGESDR1}), 
the SkyMapper Southern Survey DR2 (\citealt{SMSS}),
the Panoramic Survey Telescope and Rapid Response System (\citealt{PS1_2022}; \citealt{PS1}),
the Javalambre Photometric Local Universe Survey (\citealt{JPLUS_Sanjuan} and \citealt{JPLUS}) 
and the Southern Photometric Local Universe Survey (\citealt{SPLUS}).
The aforementioned datasets have been (re-)calibrated to a typical precision of a few mmag by using current Gaia photometry and XP spectra.
To pursuit a better calibration precision up to 0.1 per cent, it is crucial to ensure spatial uniformity of these two Gaia datasets.

In this work, we aim to perform a spatial uniformity check of Gaia DR3 photometry and XP spectra.
Given the much higher spatial uniformity of Gaia data than all other surveys, we first do a cross-validation with each other.
Upon comparing Gaia DR3 photometry with the synthetic photometry derived from its XP spectra (XPSP) in the $BP$, $RP$ and $G$ bands, we observe a spatially varying discrepancy between the two datasets, which appears to be associated with the Gaia's scanning law.
It is worth noting that BP/RP spectra and integrated BP/RP photometry are not independent, as both are obtained from the same observations. Several calibrations are applied before the integrated fluxes are produced, including corrections for bias and non-uniformities, background, and geometric calibration. Consequently, spatial patterns in the sky comparisons between BP and RP integrated and synthetic photometry are primarily linked to the final stage of the calibration process. Therefore, the small dispersion in the discrepancies is somewhat less relevant for the purposes of this paper.

We further compare Gaia DR3 datasets with the Dark Energy Survey (DES; \citealt{DES}) DR2 photometry (\citealt{DESDR2}), which currently has the best calibration precision in ground-based observations. 
As \cite{FGCM} mentioned, the calibration process of DES is entirely independent of Gaia , providing opportunity to scrutinize the origins of discrepancies between Gaia DR3 photometry and XP spectra.

This paper is organized as follows.
In Section\,2, we introduce the datasets used.
In Section\,3, we describe the method and result.
The conclusions are given in Section\,4.

\section{DATASETS} 
\subsection{ {\it Gaia} DR3 photometry and XP spectra} 

{\it Gaia} is a satellite  mission of the European Space Agency.
It focuses on astrometry, photometry, and spectroscopy of objects in the Milky Way and Local Group (\citealt{GaiaMission2016a}). 
Gaia DR3 has provided the photometry of over 1.8 billion sources in the $BP$, $RP$ and $G$ bands, as well as approximately 220 million XP spectra covering the wavelength range of $336-1020$\,nm (\citealt{dr3content}; \citealt{GaiaXPVad}; \citealt{dr3intcali}; \citealt{GaiaEDR3vali}).

The photometric calibration method employed by Gaia (\citealt{GaiaCali}) is a self-calibration approach (\citealt{GaiaCali2006}) similar to the Ubercalibration method (\citealt{uber}). This approach is based on the assumption that the physical magnitudes of non-varying sources should remain constant under varying observational conditions. The complex calibration model, encompassing numerous parameters, is effectively well-constrained through a substantial volume of repeated scanning observations.
Therefore, Gaia has enabled to achieve a superior calibration precision, with values of  3, 1.8 and 2.4 mmag in the $BP$, $RP$ and $G$ bands of Gaia EDR3 (single observation), respectively, as mentioned in \cite{GaiaEDR3vali}. 

Gaia XP spectra take the form of a projection onto 55 orthonormal Hermite functions as base-functions for both the BP and RP spectra, rather than the flux as a function of wavelength (\citealt{dr3intcali}; \citealt{dr3extcali}). Therefore, each XP spectrum consists of 110 coefficients. For our work, we transform the coefficients by the \texttt{GaiaXPy} package (\citealt{GaiaXPy}) into absolute sampled spectra, in wavelength space ranging from $336-1020$\,nm in increments of $2$\,nm.

\subsection{DES DR2} 
DES is a ground-based imaging survey in the visible and near-infrared ($grizY$ bands), covering the south Galactic cap with approximately 5,000 square degree. The DES DR2 contains all six years of DES wide area survey observations and all five years of DES supernova survey observations, which has achieved precise photometry to a depth of 24 mag at signal to noise ratio = 10 (\citealt{DESDR2}). The photometric calibration uniformity has been achieved at a level of 3 mmag in DES DR2. Its calibration process relies on the Forward Global Calibration Method (\citealt{FGCM}), which is entirely independent of Gaia.

\section{Method and Result}
\subsection{Comparison between Gaia DR3 photometry and the synthetic photometry of XP spectra}

We utilize all the 219,197,643 Gaia XP spectra to perform synthetic photometry by employing the Gaia DR3 transmission curves for the $BP$, $RP$, and $G$ bands, along with the Gaia XP spectra in the wavelength space.  
Note that the transmission curves of the $G$ and $BP$ bands slightly exceed the wavelength range of the XP spectra at the blue end, and the $G$ and $RP$ bands' transmission curves slightly exceed the wavelength range at the red end. To address this, at the blue end, we perform linear interpolation using the XP spectra fluxes in the $336-355$\,nm range to estimate the fluxes in the $320-335$\,nm range. Additionally, at the red end, we supplement the $1021-1100$\,nm fluxes with an average of the $1011-1020$\,nm fluxes. 

Based on a test using over 0.5 million sources from a region located at $R.A. = 45^\circ, Dec. = -20^\circ$ and covering about 800 square degrees, our synthetic photometry results align well with those obtained using the  \texttt{GaiaXPy} package (\citealt{GaiaXPy}). 
The overall median values of differences between our results and \texttt{GaiaXPy} are $-$0.12, 0.26, and 0.07 mmag for the $BP$, $RP$, and $G$ bands, respectively. And the median values do not show any trend with brightness. The variations in median values at different brightness levels remain within 0.1 mmag for all three bands relative to the overall median values.
Additionally, the median values of differences exhibit a very weak trend with the $BP-RP$ color, which could possibly originate from the differences in the extrapolation approaches between our results and those of \texttt{GaiaXPy}. For most sources with $0.5 < BP-RP < 2$, the variations of the median values relative to the overall median values are smaller than 0.15 mmag for all three bands.
The dispersion of the differences for all three bands increases with decreasing brightness. The typical dispersion values are 0.08, 0.16, and 0.05 mmag at $G = 12$ and 0.51, 0.89, and 0.34 mmag at $G = 16$ for the $BP$, $RP$, and $G$ bands, respectively. No spatial patterns are observed in this region for any of the three bands. As demonstrated in the following discussion, this region includes obvious spatial patterns between the photometry and spectra of Gaia DR3 and DES photometry. For the remainder of this paper, we adopt the synthetic photometry derived from absolute XP spectra and transmission curves for the $BP$, $RP$, and $G$ bands.

\begin{figure*}[ht]
\includegraphics[width=180mm]{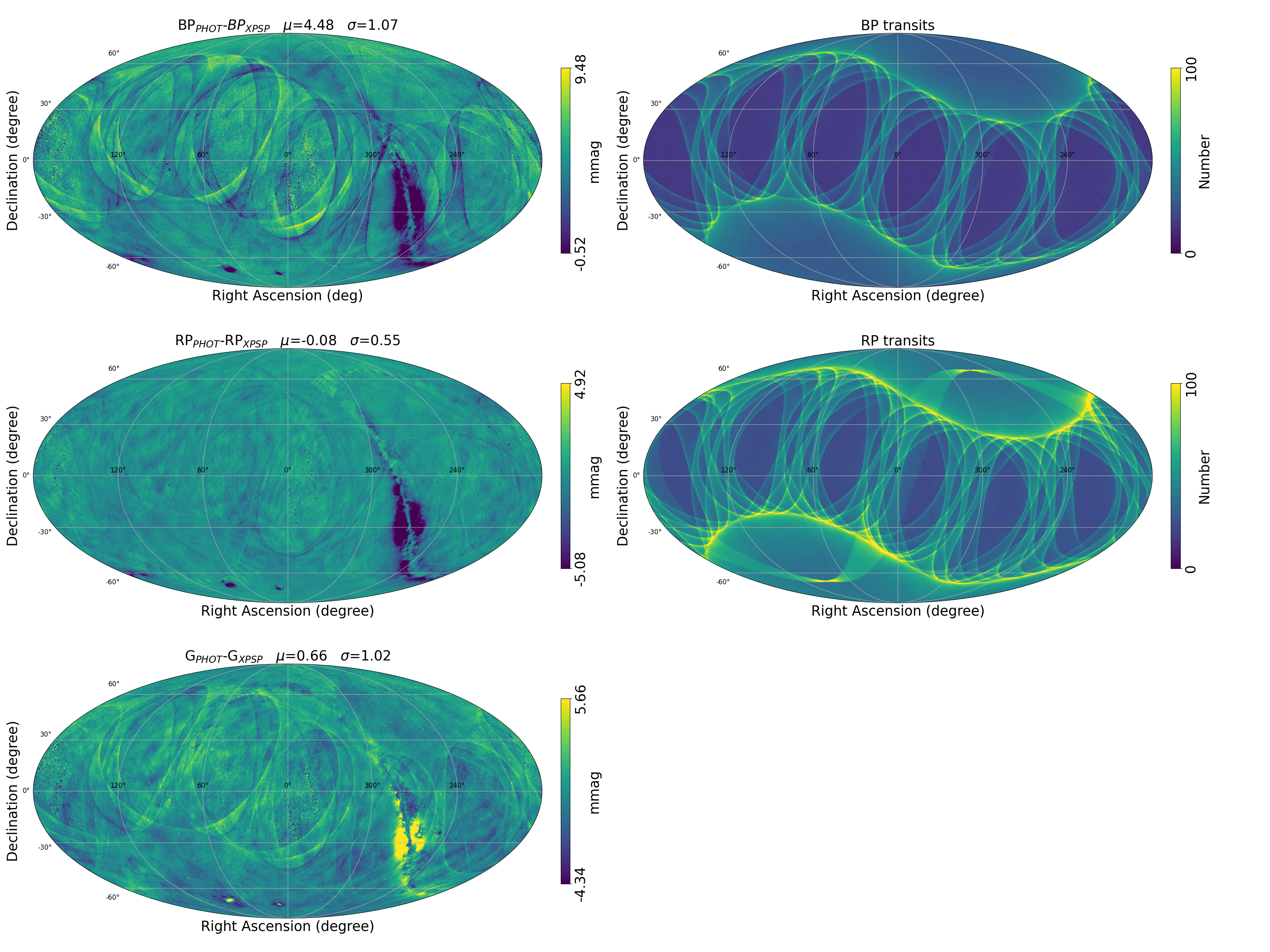}
\caption{The left panels show comparison between Gaia DR3 photometry and the synthetic photometry of XP spectra in the $BP$, $RP$ and $G$ bands. The colorbars have a range of $\pm$ 5 mmag from the median. The right panels indicate the number of transits contributing to the $BP$ and $RP$ spectra.} 
\label{Fig1}
\end{figure*}

We subsequently partition the celestial sphere into 196,608 portion by applying the HEALPix scheme (nside = 128, corresponding to a spatial resolution of $\approx$ 27.'5). 
The median discrepancies between Gaia DR3 photometry and synthetic photometry of XP spectra in the $BP$, $RP$, and $G$ bands for each portion are depicted in the left panels of Figure \ref{Fig1}. The numbers of transits contributing to the BP and RP spectra are shown in the right panels to indicate the Gaia's scanning law. One can see that the spatial discrepancy patterns are correlated with the Gaia's scanning law. The patterns are more significant in the $G$ and $BP$ bands than the $RP$ band. It is noteworthy that the spatial patterns in the $BP$ and $G$ bands exhibit similarity, with the former being approximately 1.4 times the values of the latter. Further discussions can be found in Section 3.2. Note that the BP spectra have been generated using only a fraction of the available observations to avoid periods of severe contamination, as shown in the right panels of Figure \ref{Fig1}, where the number of transits contributing to BP is smaller than that for RP. This reduction in the number of observations contributing to the synthetic BP and G may increase the influence of outliers, if outliers are the cause of the observed patterns.

We perform Gaussian fits to distributions of discrepancies, with dispersions approximately 1 mmag in the $BP$ and $G$ bands, and 0.55 mmag in the $RP$ band. It is worth mentioning that the fit does not account for the extended tail of the histogram distribution, which represents the high source density regions. In the very dense regions (e.g., Galactic center, Magellanic Clouds), Gaia photometry is strongly affected by background and crowding issues, which is related to sky subtraction. 
The different discrepancies between the $G$ band and $BP/RP$ band in very dense regions is attributed to the methodological difference in photometry, where $BP$ and $RP$ magnitudes are derived from aperture photometry, while $G$ magnitude relies on Point-Spread Function (PSF) Fitting photometry. 
Furthermore, we do not attribute the discrepancies in regions near the Galactic Plane to sources with redder colors due to high extinction, given that similar discrepancies are observed in the Magellanic Clouds region, where extinction is not particularly high.

\subsection{Comparison between Gaia DR3 and DES DR2}
To further investigate the spatial discrepancies between Gaia DR3 photometry and XP spectra, we conduct an independent test using DES DR2 photometric data. 

There are about 5.4 million common sources between Gaia DR3 XP spectra and DES DR2, using a matching radius of 1\arcsec. We apply the following selection criteria to ensure the data quality and facilitate the implementation of color-color transformation.
\begin{enumerate}
    \item $15.5<G<17$
    \item $0.5<BP-RP<1.5$
    \item $bp\_rp\_excess\_factor< (BP-RP)*0.09+1.15$
    \item $Flag<4$ for DES $grizY$ bands
    \item $e\_gmag < 0.003$, $e\_rmag < 0.003$, $i\_gmag < 0.004$, $z\_gmag < 0.005$ and $e\_Ymag < 0.01$
    \item $ExtClsCoad = 0$
\end{enumerate} 
After the above constraints, there are approximately 1.1 million common sources left. We then perform color-color transformations to convert DES $g$, $r$, $i$ and $z$ magnitudes into Gaia $BP$, $RP$, and $G$ magnitudes using third-order polynomials. The specific forms along with their corresponding coefficients and the dispersion of residuals can be found in Table \ref{Tab1}. The transformation residuals exhibit only a very weak relationship with color, up to 2 mmag. We clip outliers when calculating the dispersion of residuals and do not delve into the reasons for outliers, as they will not affect the test of spatial discrepancies.

\begin{table*}
\centering
\caption{Color-color transformation coefficients}
\label{Tab1} 
\begin{tabular}{rrrrrrrr} \hline\hline
 $Color_1$ & $Color_2$ & $a_3$ &$a_2$ & $a_1$ & $a_0$ & Residuals (mmag)\\ \hline
$g-r$ & $BP_{PHOT} - g$ & $-$0.021& $-$0.147& $-$0.122& $+$0.020& 7.78& \\
$g-r$ & $BP_{XPSP} - g$ & $-$0.022& $-$0.140& $-$0.130& $+$0.019& 7.37& \\
$r-z$ & $RP_{PHOT} - i$ & $+$0.267& $-$0.293& $+$0.137& $-$0.363& 6.32& \\
$r-z$ & $RP_{XPSP} - i$ & $+$0.267& $-$0.292& $+$0.136& $-$0.361& 6.21& \\
$g-i$ & $G_{PHOT} - r$ & $-$0.020& $-$0.046& $+$0.225& $-$0.070& 8.58& \\
$g-i$ & $G_{XPSP} - r$ & $-$0.025& $-$0.032& $+$0.213& $-$0.065& 9.05& \\
\hline
\end{tabular}
$Color_2 = a_3 \times Color_1^3 + a_2 \times Color_1^2 + a_1 \times Color_1 +a_0 $
\end{table*}

\begin{figure*}[ht]
\includegraphics[width=180mm]{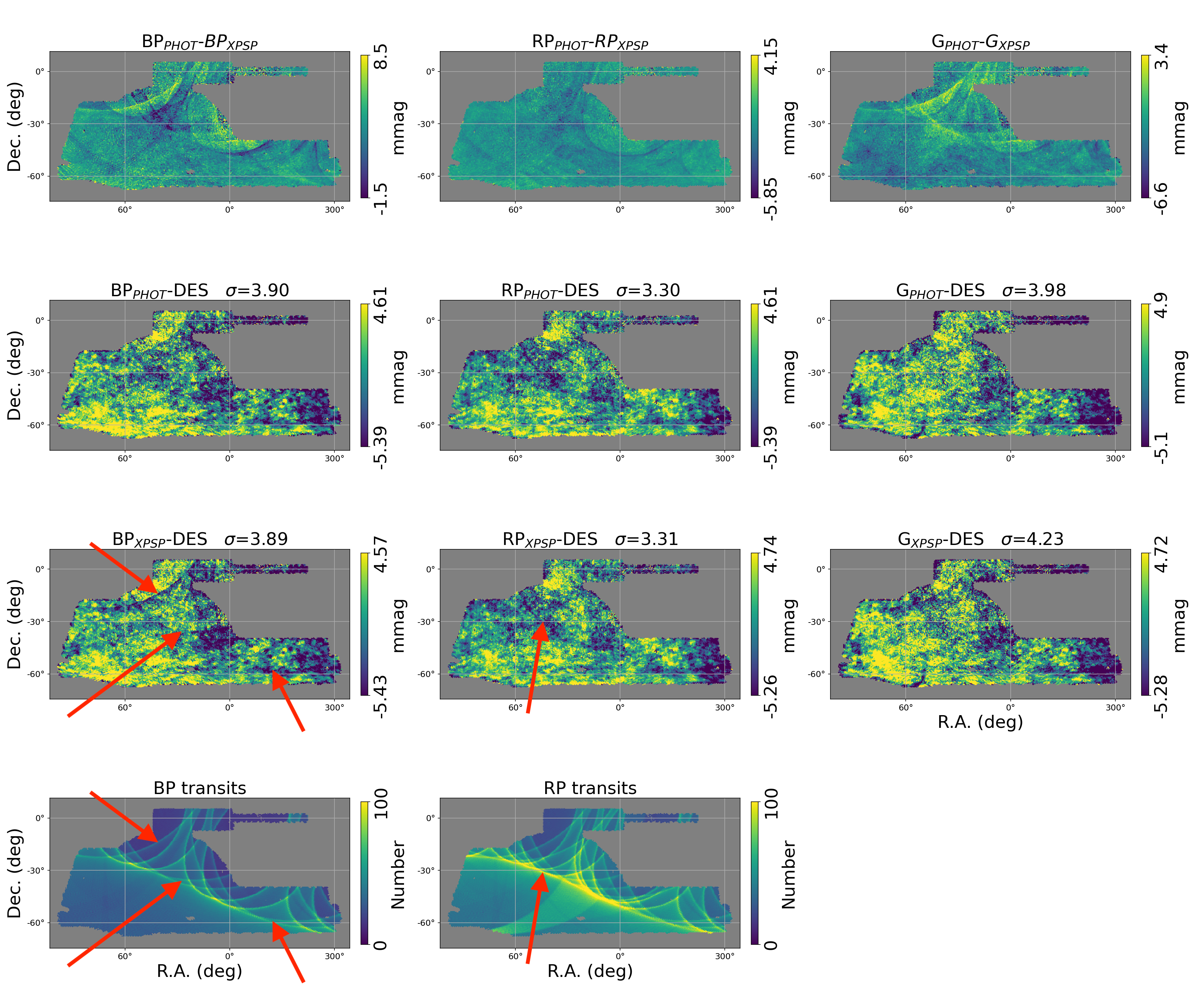}
\caption{Pairwise comparisons between Gaia DR3 photometry, the synthetic photometry of XP spectra and the DES DR2 photometry in the $BP$ (left panels), $RP$ (middle panels), and $G$ (right panels) bands. The colorbars have a range of $\pm$ 5 mmag from the median. The bottom two panels indicate the number of transits contributing to the $BP$ and $RP$ spectra. Red arrows mark the Gaia's scanning law like patterns. } 
\label{Fig3}
\end{figure*}

As in the previous section, we partition the DES footprint using the HEALPix scheme with nside of 128. The pairwise comparisons between Gaia DR3 photometry, synthetic photometry of XP spectra, and DES DR2 in the $BP$, $RP$, and $G$ bands are illustrated in Figure \ref{Fig3}. 
The first row of Figure\,\ref{Fig3} is identical to the left column of Figure\,\ref{Fig1}, with a focus on the DES footprint, providing a reference for the spatial patterns associated with Gaia's scanning law. The second row of Figure\,\ref{Fig3} illustrates the spatial discrepancies between the Gaia photometry and the DES, and the third row shows the spatial discrepancies between the synthetic photometry derived from XP spectra and the DES. And the labels give the standard deviation figure for each comparison.
The forth row is similarly identical to the right column of Figure\,\ref{Fig1}.

It can be observed that both Gaia photometry and XP spectra exhibit two different types of spatial patterns in comparison with DES, as shown in the second and third rows. The first type is the more obvious, less-structured pattern, which has no correlation with Gaia's scanning law. The origin of this less-structured pattern cannot be determined without reliable third-party high-uniformity data for comparison. Given the pattern of Gaia's scanning law, we are inclined to believe that the less-structured pattern most likely originates from DES. The second type of pattern, which is the focus of this paper, resembles Gaia's scanning law. This structure manifests as arc patterns in the second and third rows, specifically identified in the third row with red arrows. In the fourth row, the corresponding Gaia's scanning law patterns are marked with same red arrows in the corresponding positions.
The three arcs marked in the $BP_{XPSP}-DES$ panel are also observed in the $BP_{PHOT}-DES$ and $G_{XPSP}-DES$ panels, although less clearly. Similarly, the arc marked in the $RP_{XPSP}-DES$ panel is observed in the $RP_{PHOT}-DES$ and $G_{XPSP}-DES$ panels, but it is less clear in the $G_{XPSP}-DES$ panel.
Due to the overlapping passbands of both $BP$ and $G$ with the wavelength range of the $BP$ spectra, and of $RP$ and $G$ with the wavelength range of the $RP$ spectra, any effect present in the $BP$ spectra will automatically propagate into the synthetic $BP$ and $G$ photometry, and any effect present in the $RP$ spectra will automatically propagate into the synthetic $RP$ and $G$ photometry. This propagation occurs at a somewhat reduced level due to the broader coverage of the $G$ passband compared to that of $BP$ and $RP$.
Notably only in the $G_{PHOT}-DES$ panel are Gaia's scanning law-like spatial patterns less prominent. This is a significant improvement in Gaia DR3 $G$ band photometry compared to Gaia DR2 (\citealt{DESDR2}).
Therefore, combined with the previously discussed results, our findings indicate potential issues in both the $BP$ and $RP$ photometry and spectra, with the $BP$ spectra exhibiting stronger spatial structures. In contrast, the $G$ photometry demonstrates relatively better spatial uniformity.

The dispersion observed in the comparison between $G_{PHOT}$ and DES is greater than that reported in \cite{Rykoff2023}. Two possible reasons could explain this issue.
Firstly, to compare both Gaia photometry and synthetic photometry with the DES simultaneously, we require the same sources to be used in Figure \ref{Fig3}. Due to the brightness limitation of Gaia BP/RP spectra (G $<$ 17.65), significantly fewer sources are used compared to the results in \cite{Rykoff2023}, where the limit is G $<$ 20. Secondly, only one color is used in the transformations between DES filters and the Gaia bands, thus neglecting the influence of metallicities and reddening, which could result in a greater dispersion of residual for individual sources in the transformation.
However, the increased dispersion caused by the aforementioned processes will not impact the conclusion regarding the existence of a spatial pattern akin to Gaia’s scanning law.

\section{Conclusion}
In this work, we evaluate the spatial uniformity of Gaia DR3 photometry and XP spectra by a comparison between the photometry in the $BP$, $RP$, and $G$ bands and their corresponding synthetic photometry derived from the XP spectra. Our results reveal minimal dispersion in the discrepancies, quantified as 1.07 mmag for the $BP$ band, 0.55 mmag for the $RP$ band, and 1.02 mmag for the $G$ band. Nonetheless, the discrepancies manifest obvious spatial patterns aligned with Gaia's scanning law. The observed patterns exhibit notable similarities in the $BP$ and $G$ bands, and significantly surpasses those in the $RP$ band.

We subsequently conduct an independent test using DES DR2 photometry, which reveals Gaia's scanning law like patterns in both Gaia DR3 photometry and XP spectra, particularly in the $BP$ and $G$ bands. Additionally, the test uncovers less-structured discrepancies between the Gaia and DES data. These findings further suggest that the observed spatial patterns are more likely attributable to the BP spectra.

Our result corroborates the high spatial uniformity of Gaia data at the mmag level and a significant enhancement in the spatial uniformity of the Gaia DR3 photometry relative to the Gaia DR2 photometry (\citealt{DESDR2}), as well as suggests that the spatial uniformity of Gaia DR3 XP spectra could be further improved in the future.

\vspace{7mm} \noindent {\bf Acknowledgments}

{The authors thank the referee for the suggestions that improved the quality of this work. This work is supported by the National Natural Science Foundation of China through the project NSFC 12222301, 12173007, and NSFC 11603002, the National Key Basic R\&D Program of China via 2019YFA0405503. 
We acknowledge the science research grants from the China Manned Space Project with NO. CMS-CSST-2021-A08 and CMS-CSST-2021-A09.
This work has made use of data from the European Space Agency (ESA) mission {\it Gaia} (\url{https://www.cosmos.esa.int/gaia}), processed by the Gaia Data Processing and Analysis Consortium (DPAC, \url{https:// www.cosmos.esa.int/web/gaia/dpac/ consortium}). Funding for the DPAC has been provided by national institutions, in particular the institutions participating in the Gaia Multilateral Agreement. 
}

{}
\end{CJK*}

\begin{thebibliography}{}
\bibitem[Abbott et al.(2021)]{DESDR2} Abbott, T.~M.~C., Adam{\'o}w, M., Aguena, M., et al.\ 2021, \apjs, 255, 20. doi:10.3847/1538-4365/ac00b3
\bibitem[Burke et al.(2018)]{FGCM} Burke, D.~L., Rykoff, E.~S., Allam, S., et al.\ 2018, \aj, 155, 41. doi:10.3847/1538-3881/aa9f22
\bibitem[Carrasco et al.(2016)]{GaiaCali} Carrasco, J.~M., Evans, D.~W., Montegriffo, P., et al.\ 2016, \aap, 595, A7. doi:10.1051/0004-6361/201629235
\bibitem[Carrasco et al.(2021)]{dr3intcali} Carrasco, J.~M., Weiler, M., Jordi, C., et al.\ 2021, \aap, 652, A86. doi:10.1051/0004-6361/202141249
\bibitem[Dark Energy Survey Collaboration et al.(2016)]{DES} Dark Energy Survey Collaboration, Abbott, T., Abdalla, F.~B., et al.\ 2016, \mnras, 460, 1270. doi:10.1093/mnras/stw641
\bibitem[De Angeli et al.(2022)]{GaiaXPVad} De Angeli, F., Weiler, M., Montegriffo, P., et al.\ 2022, arXiv:2206.06143. doi:10.48550/arXiv.2206.06143
\bibitem[Fan et al.(2023)]{SAGESDR1} Fan, Z., Zhao, G., Wang, W., et al.\ 2023, \apjs, 268, 9. doi:10.3847/1538-4365/ace04a


\bibitem[Gaia Collaboration et al.(2016)]{GaiaMission2016a} Gaia Collaboration, Prusti, T., de Bruijne, J.~H.~J., et al.\ 2016, \aap, 595, A1. doi:10.1051/0004-6361/201629272
\bibitem[Gaia Collaboration et al.(2022)]{dr3content} Gaia Collaboration, Vallenari, A., Brown, A.~G.~A., et al.\ 2022, arXiv:2208.00211. doi:10.48550/arXiv.2208.00211
\bibitem[Huang et al.(2021)]{SMSS} Huang, Y., Yuan, H., Li, C., et al.\ 2021, \apj, 907, 68. doi:10.3847/1538-4357/abca37
\bibitem[Huang \& Yuan(2022)]{S82} Huang, B. \& Yuan, H.\ 2022, \apjs, 259, 26. doi:10.3847/1538-4365/ac470d
\bibitem[Huang et al.(2024)]{HuangXPcor} Huang, B., Yuan, H., Xiang, M., et al.\ 2024, \apjs, 271, 13. doi:10.3847/1538-4365/ad18b1
\bibitem[Jordi et al.(2006)]{GaiaCali2006} Jordi, C., H{\o}g, E., Brown, A.~G.~A., et al.\ 2006, \mnras, 367, 290. doi:10.1111/j.1365-2966.2005.09944.x



\bibitem[L{\'o}pez-Sanjuan et al.(2023)]{JPLUS_Sanjuan} L{\'o}pez-Sanjuan, C., V{\'a}zquez Rami{\'o}, H., Xiao, K., et al.\ 2023, arXiv:2301.12395. doi:10.48550/arXiv.2301.12395
\bibitem[Montegriffo et al.(2022)]{dr3extcali} Montegriffo, P., De Angeli, F., Andrae, R., et al.\ 2022, arXiv:2206.06205. doi:10.48550/arXiv.2206.06205
\bibitem[Niu et al.(2021)]{NZXGaiaColor} Niu, Z., Yuan, H., \& Liu, J.\ 2021, \apjl, 908, L14. doi:10.3847/2041-8213/abe1c2
\bibitem[Yang et al.(2021)]{YangGaiaMag} Yang, L., Yuan, H., Zhang, R., et al.\ 2021, \apjl, 908, L24. doi:10.3847/2041-8213/abdbae




\bibitem[Padmanabhan et al.(2008)]{uber} Padmanabhan, N., Schlegel, D.~J., Finkbeiner, D.~P., et al.\ 2008, \apj, 674, 1217. doi:10.1086/524677

\bibitem[Riello et al.(2021)]{GaiaEDR3vali} Riello, M., De Angeli, F., Evans, D.~W., et al.\ 2021, \aap, 649, A3. doi:10.1051/0004-6361/202039587
\bibitem[Ruz-Mieres(2022)]{GaiaXPy} Ruz-Mieres, D.\ 2022, gaia-dpci/GaiaXPy: GaiaXPy 1.2.0, Zenodo, doi:10.5281/zenodo.7015044
\bibitem[Thanjavur et al.(2021)]{S82_2021} Thanjavur, K., Ivezi{\'c}, {\v{Z}}., Allam, S.~S., et al.\ 2021, \mnras, 505, 5941. doi:10.1093/mnras/stab1452
\bibitem[Rykoff et al.(2023)]{Rykoff2023} Rykoff, E.~S., Tucker, D.~L., Burke, D.~L., et al.\ 2023, arXiv:2305.01695. doi:10.48550/arXiv.2305.01695



\bibitem[Xiao et al.(2023)]{SAGES} Xiao, K., Yuan, H., Huang, B., et al.\ 2023, arXiv:2307.13238. doi:10.48550/arXiv.2307.13238
\bibitem[Xiao \& Yuan(2022)]{PS1_2022} Xiao, K. \& Yuan, H.\ 2022, \aj, 163, 185. doi:10.3847/1538-3881/ac540a
\bibitem[Xiao et al.(2023)]{PS1} Xiao, K., Yuan, H., Huang, B., et al.\ 2023, \apjs, 268, 53. doi:10.3847/1538-4365/acee73
\bibitem[Xiao et al.(2023)]{JPLUS} Xiao, K., Yuan, H., L{\'o}pez-Sanjuan, C., et al.\ 2023, \apjs, 269, 58. doi:10.3847/1538-4365/ad0645
\bibitem[Xiao et al.(2023)]{SPLUS} Xiao, K., Huang, Y., Yuan, H., et al.\ 2023, arXiv:2309.11533. doi:10.48550/arXiv.2309.11533














\end{thebibliography}
\end{document}